\documentclass[%
preprint,%
 amssymb, amsmath,%
 aip,cha,%
]{revtex4-1}

\usepackage{bm}%
\usepackage[colorlinks=true,linkcolor=blue]{hyperref}%
\usepackage{graphicx}
\usepackage{siunitx}
\usepackage{amsmath}
\usepackage{chngcntr}
\expandafter\ifx\csname package@font\endcsname\relax\else
 \expandafter\expandafter
 \expandafter\usepackage
 \expandafter\expandafter
 \expandafter{\csname package@font\endcsname}%
\fi
\begin{document}

\title{Cooper-pair splitting in two parallel InAs nanowires}

\author{S. Baba}
\thanks{These two authors contributed equally}
\affiliation{Department of Applied Physics, University of Tokyo, 7-3-1 Hongo, Bunkyo-ku, Tokyo 113-8656, Japan}

\author{C. J\"unger}
\thanks{These two authors contributed equally}
\email{christian.juenger@unibas.ch}
\affiliation{Department of Physics, University of Basel, Klingelbergstrasse 82, CH-4056 Basel, Switzerland}

\author{S. Matsuo}
\email{matsuo@ap.t.u-tokyo.ac.jp}
\affiliation{Department of Applied Physics, University of Tokyo, 7-3-1 Hongo, Bunkyo-ku, Tokyo 113-8656, Japan}

\author{A. Baumgartner}
\affiliation{Department of Physics, University of Basel, Klingelbergstrasse 82, CH-4056 Basel, Switzerland}

\author{Y. Sato}
\affiliation{Department of Applied Physics, University of Tokyo, 7-3-1 Hongo, Bunkyo-ku, Tokyo 113-8656, Japan}

\author{H. Kamata}
\affiliation{Department of Applied Physics, University of Tokyo, 7-3-1 Hongo, Bunkyo-ku, Tokyo 113-8656, Japan}

\author{K. Li}
\affiliation{Beijing Key Laboratory of Quantum Devices, Key Laboratory for the Physics and Chemistry of Nanodevices and Department of Electronics, Peking University, Beijing 100871, China}

\author{S. Jeppesen}
\affiliation{Division of Solid State Physics, Lund University, Box 118, SE-221 00 Lund, Sweden}

\author{L. Samuelson}
\affiliation{Division of Solid State Physics, Lund University, Box 118, SE-221 00 Lund, Sweden}

\author{ H.-Q. Xu}
\affiliation{Beijing Key Laboratory of Quantum Devices, Key Laboratory for the Physics and Chemistry of Nanodevices and Department of Electronics, Peking University, Beijing 100871, China}
\affiliation{Division of Solid State Physics, Lund University, Box 118, SE-221 00 Lund, Sweden}

\author{C. Sch{\"o}nenberger}
\email{christian.schoenenberger@unibas.ch}
\affiliation{Department of Physics, University of Basel, Klingelbergstrasse 82, CH-4056 Basel, Switzerland}

\author{S. Tarucha}
\email{tarucha@ap.t.u-tokyo.ac.jp}
\affiliation{Department of Applied Physics, University of Tokyo, 7-3-1 Hongo, Bunkyo-ku, Tokyo 113-8656, Japan}
\affiliation{Center for Emergent Matter Science, RIKEN, 2-1 Hirosawa, Wako-shi, Saitama 351-0198, Japan}


\begin{abstract}
We report on the fabrication and electrical characterization of an InAs double - nanowire (NW) device consisting of two closely placed parallel NWs coupled to a common superconducting electrode on one side and individual normal metal leads on the other. In this new type of device we detect Cooper-pair splitting (CPS) with a sizeable efficiency of correlated currents in both NWs. In contrast to earlier experiments, where CPS was realized in a single NW, demonstrating an intra\-wire electron pairing mediated by the superconductor (SC), our experiment demonstrates an \emph{inter\-wire} interaction mediated by the common SC. The latter is the key for the realization of zero-magnetic field Majorana bound states, or Parafermions; in NWs and therefore constitutes a milestone towards topological superconductivity. In addition, we observe transport resonances that occur only in the superconducting state, which we tentatively attribute to Andreev Bound states and/or Yu-Shiba resonances that form in the proximitized section of one NW.
\end{abstract}

\maketitle
\section{Introduction}
Topologically protected electronic states in nanostructures have recently attracted wide attention, as they may provide fundamental building blocks for quantum computation.\cite{Sarma2015, Hoffman2016} Recent advances in material science and device fabrication resulted in considerable progress towards the generation and detection of topologically protected bound states in topologically non-trivial semiconducting nanowires (NWs), so called  Majorana Fermions (MF).\cite{Mourik2012,Deng2012,Albrecht2016,Deng2016,Guel2018} Two MFs can combine together into a regular Fermion that is why MFs are also known as $Z_2$ Fermions. MFs are predicted to have non-Abelian braid statistics and may provide a platform for topological quantum computation.\cite{Alicea2010,Oreg2010}
However, one can not implement all required operations for universal quantum computation in quantum bits (qubits) based on MFs by using topologically protected braiding. In this respect, $Z_4$ Fermions, also known as Parafermions (PFs), are better as they allow for a larger set of operations.\cite{Pachos2012}
Recently, it has theoretically been predicted that PFs can be generated in a system based on two NWs with different spin orbit interaction coupled to a common superconducting electrode.\cite{Keselman2013,Klinovaja2014,Gaidamauskas2014} The SC induces both a pairing interaction within each NW and between the two NWs due to Cooper-pair splitting (CPS). In order to realize PFs, the interwire coupling must dominate.\cite{Klinovaja2014} It has been shown that this is possible in systems with strong electron-electron interaction, such as nanoscaled semiconducting wires.\cite{Sato2012,Bena2002,Recher2002}
Another advantage of the parallel two-wire approach, even if PFs are not formed, is the fact, that for large interwire pairing an external magnetic field is not required or only a small field is enough to reach the topological phase.\cite{Thakurathi2018} A large magnetic field is a limiting factor, because of the critical magnetic field of the SC. It is therefore crucial both for MF-based charge qubits and for the realization of PFs to demonstrate an appreciable magnitude of interwire pairing interaction mediated by a SC. This coupling is also known as crossed-Andreev reflection or CPS.\cite{Recher2001,Sato2012,Chevallier2012}
In the last few years, several CPS experiments have been performed using different platforms, mainly {\em single} NWs,\cite{Hofstetter2009,Hofstetter2011,Das2012,Fueloep2014, Fueloep2015} Carbon Nanotubes \cite{Herrmann2010,Schindele2012} and graphene.\cite{Tan2015,Borzenets2016} Splitting efficiencies close to 100\% have been reported,\cite{Schindele2012} demonstrating that intrawire pairing can exceed local Cooper pair tunneling. Up to now, all NW based CPS devices consisted of a single NW contacted by two normal metal electrodes and one superconducting contact in between. In order to assess the interwire pairing in double NW structures, it is essential to investigate CPS in such a system. In this work we demonstrate CPS in a parallel double NW device, which is an important first step towards topological quantum computation with PFs.

\section{Scheme and sample}
\begin{figure}[h]
	\begin{center}
	\includegraphics[width=0.9\columnwidth]{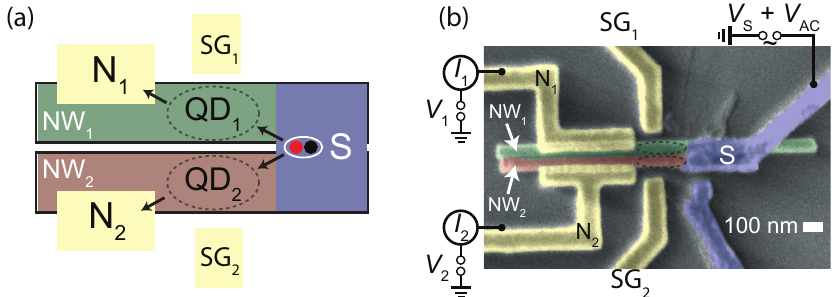}
    \end{center} 	
	\caption{(a) Schematic depiction of Cooper pair splitting in a double NW setup. Two InAs NWs, $\textrm{NW}_{1/2}$, are located in parallel, coupled by a common superconducting lead S and individual normal metal leads, $\textrm{N}_{1/2}$. $\textrm{QD}_{1/2}$ are tuned separately by local sidegates $\textrm{SG}_{1/2}$. (b) Scanning electron microscope image with false color of the device consisting of $\textrm{NW}_{1/2}$ with a common Al contact (S) on the right and individual Au contacts ($\textrm{N}_{1/2}$) on the left and sidegates, $\textrm{SG}_{1/2}$. The measurement setup is also shown.}
	\label{fig:CPS-DeviceSetup}
\end{figure}

We investigate a device shown schematically in figure \ref{fig:CPS-DeviceSetup}(a). Two InAs semiconducting NWs with large spin orbit interaction are placed in parallel ($\textrm{NW}_1$ green, $\textrm{NW}_2$ red) and electrically coupled by a common superconducting electrode $\textrm{S}$ (blue). Both NWs are contacted by individual normal metal leads $\textrm{N}_{1/2}$ (yellow). Sidegates $\textrm{SG}_{1/2}$ (yellow) are located on each side of the NWs, in order to separately tune the chemical potentials of the quantum dots (QDs), which form between $\textrm{N}_{1/2}$ and $\textrm{S}$. We note, that the exact location of the QDs is not known, since we do not use additional barrier gates to terminate the QDs.\cite{Fasth2005}
%
%
However, it is clear that both $\textrm{N}_{1/2}$ and $\textrm{S}$ induce a potential step from which (partial) electron reflection is possible and QD bound states can form. We also point out already here that the electronic boundary conditions on the $\textrm{S}$ side may change if $\textrm{S}$ is in the normal or superconducting state, due to the proximity effect. Besides local Cooper pair tunneling from S to $\textrm{N}_{1/2}$,\cite{Gramich2015}  Cooper pairs (white circle with red/black dot) can be split, resulting in a non-local current consisting of entangled single electrons. This process is expected to be large if both QDs, $\textrm{QD}_{1/2}$, are in resonance and the electrons can sequentially tunnel from the SC to the two normal metal leads.

\subsection{Fabrication and Characterization}
The InAs NWs used in this study are grown by Chemical Beam Epitaxy along the $\langle$111$\rangle$ direction. They have a diameter of about \SI{80}{\nano \meter} and possess pure Wurtzite crystal structure.  After transferring NWs from the growth chip to the substrate by standard dry transfer, we use scanning electron microscopy, to select NWs naturally lying next to each other. It is important to note that the NWs are electronically disconnected by their native oxide, which is about \SIrange{2}{3}{\nano \meter} thick surrounding each NW. Next, we deposit the common superconducting lead $\textrm{S}$ made of Ti/Al (thickness: \SI{3}{\nano \meter}/\SI{90}{\nano \meter}) after removing the native oxide at the contact area using a solution of $({NH}_{4})_2S_x$.\cite{Suyatin2007} Afterwards, the individual normal metal contacts $\textrm{N}_{1/2}$ made of Ti/Au (\SI{3}{\nano \meter}/\SI{130}{\nano \meter}) are deposited at the same time as the local sidegates $\textrm{SG}_{1/2}$. A false color scanning electron microscopy image of the device is shown in figure \ref{fig:CPS-DeviceSetup}(b). The distance between the source Al contact and the Au drain contacts is about \SI{250}{\nano \meter}.\\\\
%
All measurements were carried out in a dilution refrigerator with a base temperature of about  \SI{50}{\milli \kelvin}. Differential conductance has been measured for the respective NWs simultaneously using synchronized lock-in techniques (see figure \ref{fig:CPS-DeviceSetup}(b)). Characterization measurements (see figure \ref{fig:D_diamonds} in appendix) indicate two individual QDs $\textrm{QD}_1$ and $\textrm{QD}_2$ in each of the NWs, similar to previous measurements.\cite{Baba2017} From Coulomb blockade measurements we extract the following parameters for the two QDs for the charging energy $U$, single particle level spacing $\epsilon$ and the life-time broadening of the QD eigenstates $\Gamma$ to the leads: $U_{1,2} = 0.5-0.7$\,meV,
$\epsilon_1 = 0.3-0.5$\,meV, $\epsilon_2 = 0.1 - 0.3$\,meV,
$\Gamma_1 = 0.1-0.2$\,meV and $\Gamma_2 = 0.2 - 0.3$\,meV for $\textrm{QD}_1$ and $\textrm{QD}_2$, respectively.
Both quantum dots hold similar properties, implying that each QD is formed between the Al contact and individual Au contacts. In addition, we observe a slight suppression of conductance for some regions within the superconducting energy gap $\delta$ of about \SI{150}{\micro \eV}, which is similar to other experiments, see appendix.\cite{Das2012}
We note here, that since $\Delta < \Gamma$, local pair tunneling should exceed CPS.\cite{Recher2012,Schindele2012}
We also emphasize that we cannot distinguish the individual tunnel coupling strengths of each QD to either $\textrm{S}$ or $\textrm{N}$. The respective tunnel-rate ratio has an important effect on the magnitude of CPS. In particular, CPS can appear to be suppressed in the experiment if tunneling out of the QD into the drain electrode is the rate-limiting step.\cite{Schindele2012,Fueloep2015}

\section{Cooper pair Splitting in double NW}
\begin{figure}[b]
	\begin{center}
	\includegraphics[width=0.9\columnwidth]{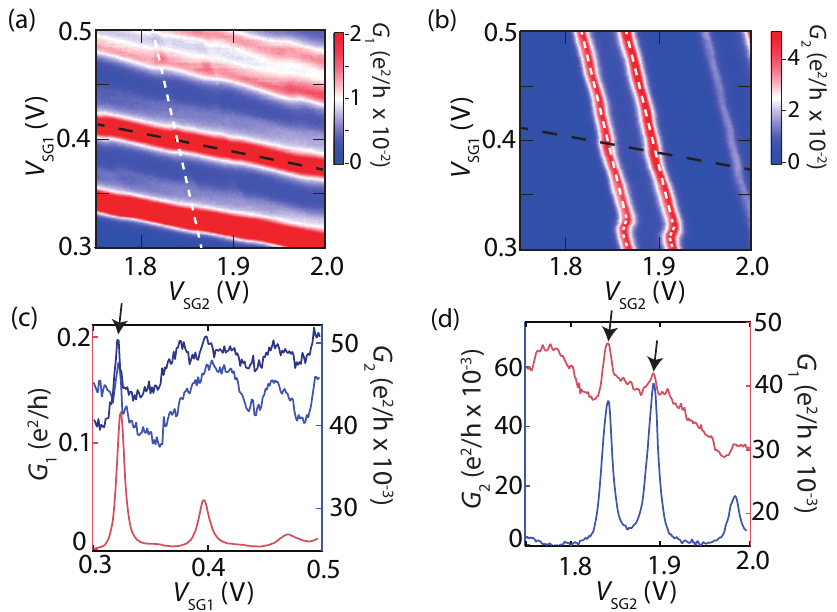}
    \end{center} 	
	\caption{(a) Differential conductance $G_1$ of $\textrm{QD}_1$ as a function of $V_{SG1}$ and $V_{SG2}$ and (b)  $G_2$ respectively for $\textrm{QD}_2$. (c) cross sections along dashed white lines of (a) and (b). (d) cross section along black dashed line of (a) and (b).}
	\label{fig:CPS-gate}
\end{figure}
In figure \ref{fig:CPS-gate}(a) and \ref{fig:CPS-gate}(b) the simultaneously measured differential conductance  $G_1$ through $\textrm{QD}_1$ and $G_2$ through $\textrm{QD}_2$ are shown, both as a function of the side-gate voltages $V_{SG1}$ and $V_{SG2}$. These measurement are done at zero bias and without an external magnetic field. Varying the sidegate voltage $V_{SG1}$ tunes $\textrm{QD}_1$ through several Coulomb blockade resonances, resulting in conductance peaks in $G_1$. Similarly  $V_{SG2}$ tunes the resonances of $\textrm{QD}_2$ in $\textrm{NW}_2$. Each resonance signifies a change in charge state of the respective QD. The charge on one QD can be sensed by the other QD, due to the capacitive coupling between $\textrm{QD}_1$ and $\textrm{QD}_2$. In our experiment, $\textrm{QD}_2$ acts as a good sensor for the charge on $\textrm{QD}_1$, as the Coulomb blockade resonance lines of $\textrm{QD}_1$ shift whenever the charge on $\textrm{QD}_2$ changes by one electron, see Fig.\ref{fig:CPS-gate}(b). Due to capacitive crosstalk from $V_{SG1}$ on $\textrm{QD}_2$ ($V_{SG2}$ on $\textrm{QD}_1$ respectively) the resonance positions are slightly tilted in both graphs.
At certain gate voltages, when both QDs are in resonance, an increase of conductance can be observed on both sides. This can be seen more clearly in the cross sections indicated by black and white dashed lines Fig.\ref{fig:CPS-gate}(a,b).
We observe an enhancement of $\textrm{G}_1$ along the black dashed line in Fig.\ref{fig:CPS-gate}(a) at the peak positions of $\textrm{G}_2$ for the two resonances at $V_{SG2}\approx 1.87$\,V and $V_{SG2}\approx 1.91$\,V (see arrows), while other possible correlations are less clear and disappear in the background noise. A similar characteristics can be found in cross sections of $\textrm{G}_2$ along the white dashed line in  Fig.\ref{fig:CPS-gate}(b) at peak positions of $\textrm{G}_1$. Here, the positive correlation is very clear at $V_{SG1}\approx 0.32$\,V (arrow), while there is only a weak quite broadened correlation visible for the other two resonances at $V_{SG1}\approx 0.4$\,V and $V_{SG1}\approx 0.49$\,V.
Hence, we observe clear positive correlation between $\textrm{G}_1$ and $\textrm{G}_2$ on three resonances, which we assign to CPS from $\textrm{S}$ into $\textrm{QD}_1$ and $\textrm{QD}_2$. We repeat the same measurement in the absence of superconductivity by applying an out of plane magnetic field of \SI{250}{\milli \tesla}, which is larger than the critical field of the Al contact. In this case the previously positive correlations between $\textrm{G}_1$ and $\textrm{G}_2$ disappear fully, proofing that the positive correlation in conductance originates from CPS (see figure \ref{fig:E_CPS_Bfield} in appendix). We define the CPS efficiency as 2$\textrm{G}_{CPS}$/$\textrm{G}_{total}$ resulting in a maximum efficiency of $\approx$ 20\%, similar to the largest reported values on single NW devices.

\begin{figure}[b]
	\begin{center}
	\includegraphics[width=0.9\columnwidth]{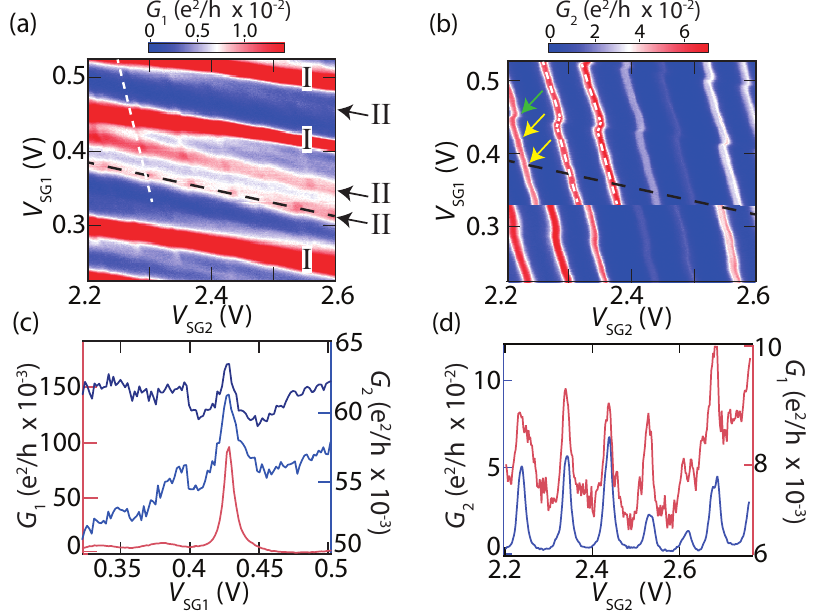}
    \end{center} 	
	\caption{(a) Differential conductance $G_1$ of $\textrm{QD}_1$ as a function of $V_{SG1}$ and $V_{SG2}$, second set of resonances are denoted with ${\rm II}$ (black arrows). (b)  Differential Conductance $G_2$ respectively for $\textrm{QD}_2$. (c) cross sections along dashed white lines of (a) and (b). (d) cross section along black dashed line of (a) and (b).}
	\label{fig:B_2nCPS}
\end{figure}

In the following we investigate the same type of measurement as discussed in figure \ref{fig:CPS-gate}, for a different gate voltage region. The data is shown in figure \ref{fig:B_2nCPS}. Here, we observe two sets of resonances for $\textrm{QD}_1$. Besides the QD levels, we already discussed in figure \ref{fig:CPS-gate} (indicated with ${\rm I}$), we detect a second set of resonances (referred to ${\rm II}$). For $\textrm{QD}_2$ we observe only one set of QD resonances. The resonances denoted with ${\rm II}$ are significantly different from the one denoted with ${\rm I}$.
First, the amplitude of type ${\rm II}$ resonances is only half of the value of type ${\rm I}$. Furthermore, type ${\rm II}$ have a different slope than type ${\rm I}$, indicating different capacitive coupling ratio between $\textrm{SG}_1$ and $\textrm{SG}_2$. The broadening differs roughly by a factor of two: $\Gamma_{\rm I}$= \SI{0.5}{\meV} whereas $\Gamma_{\rm II}$= \SI{1.2}{\meV}. Most strikingly, the resonances of type ${\rm II}$ vanish completely when an external magnetic field is applied, i.e. these resonances are only present in the super\-conducting state (not shown). In addition, $\textrm{QD}_{2}$ only reacts on a change in charge state of $\textrm{QD}_{1}$ for resonances of type ${\rm I}$ (e.g. green arrow in \ref{fig:B_2nCPS}(b)). For the resonances of type ${\rm II}$, there seems to be no change in the charge state, as $\textrm{QD}_{2}$ does not sense these resonances (e.g. yellow arrows in \ref{fig:B_2nCPS}(b)). We therefore conclude that the resonance lines of type ${\rm II}$ are not Coulomb blockade resonances. They must have their origin within the superconducting phase, most likely near $\textrm{S}$. Though these features are not Coulomb blockade resonances, they can still be used to test CPS by conductance correlations. We observe a clear positive correlation between the conductances $\textrm{G}_1$ and $\textrm{G}_2$ in figure \ref{fig:B_2nCPS}(d) as well as in \ref{fig:B_2nCPS}(c) for both types of resonances. We estimate a splitting efficiency of about $\approx$13\% for type ${\rm II}$, which is of similar magnitude as the one obtained for resonances of type ${\rm II}$ in the gate region of Fig.\ref{fig:CPS-gate}.\\

The measurements of type ${\rm II}$ resonances suggest the existence of sub-gap states which are not located in $\textrm{QD}_{1}$, but rather in the lead connecting to $\textrm{S}$. Since these states are gate-tunable, they are not fully screened by $\textrm{S}$.
We therefore propose that a proximitized region is formed in NW1 that extends to some distance out from $\textrm{S}$. $\textrm{QD}_{1}$ is coupled to this proximitized lead. Within the lead, bound states can form due to potential fluctuations and residual disorder. There are two kinds of bound states, Andreev bound states (ABS) \cite{Pillet2010,Dirks2011,Gramich2017} or Yu-Shiba Rusinov (YSR)\cite{Jellinggaard2016} states. These states do not usually occur at zero energy, but can be tuned electrically to zero energy, signaling a ground state transition between the proximitized lead region and the bulk of the SC. This gives in effect rise to a density-of-state peak in the gap of the SC, enhancing the subgap conductance which we measure. In this case the two electrons that are launched by CPS are transmitted in a different way to the respective drain electrodes. The electron that takes the path through $\textrm{QD}_{2}$ is transferred by the usual (resonant) sequential tunneling, while the one that takes the path through $\textrm{QD}_{1}$ is transferred by co-tunneling. One might expect that this suppresses CPS as the latter process corresponds to a low probability for out-tunneling into the drain contact. However, due to the sub-gap resonance in the proximitized lead, this process is enhanced and one can therefore reach almost similar CPS efficiencies.
\section{Conclusion}
In summary, we demonstrate the fabrication of an electronic device, consisting of two closely placed parallel InAs NWs, contacted by a common superconducting lead and individual normal metal leads. By addressing individual sidegates we can separately tune the QDs formed in each NW. When both QDs are in resonance, we observe CPS with efficiencies up to 20\%. For certain gate voltages, we detect a second set of QD resonances in one of the NWs, which only appears in the superconducting state. The second set of resonances do not correspond to Coulomb blockade resonances, hence, are not related to a change in the charge state of the QD. Since they only appear in the superconducting state, we tentatively assign the second set to subgap states in the lead that connects to the SC, which are most likely caused by superconducting bound states (Andreev and or Yu-Shiba Rusinov states). For the first time we provide a platform, suitable to implement the next milestone in topological quantum computation, namely Parafermions.

\section{Acknowledgements}
  This work was partially supported by a Grant-in-Aid for Young Scientific Research (A) (Grant No. JP15H05407), Grant-in-Aid for Scientific Research (A) (Grant No. JP16H02204), Grant-in-Aid for Scientific Research (S) (Grant No. JP26220710), JSPS Research Fellowship for Young Scientists (Grant No. JP14J10600), and the JSPS Program for Leading Graduate Schools (MERIT) from JSPS, Grants-in-Aid for Scientific Research on Innovative Area “Nano Spin Conversion Science” (Grants No. JP15H01012 and No. JP17H05177) and a Grant-in-Aid for Scientific Research on Innovative Area “Topological Materials Science” (Grant No. JP16H00984) from MEXT, JST CREST (Grant No. JPMJCR15N2), Murata Science Foundation and the ImPACT Program of Council for Science, Technology and Innovation (Cabinet Office, Government of Japan). Part of this research was performed within the Nanometer Structure Consortium/NanoLund-environment, using the facilities of Lund Nano Lab, with support from the Swedish Research Council (VR), the Swedish Foundation for Strategic Research (SSF) and from Knut and Alice Wallenberg Foundation (KAW). It was also supported by the Swiss National Science Foundation, the Swiss Nanoscience Institute (SNI), the NCCR on Quantum Science and Technology and the H2020 project QuantERA.
\newpage
\section*{Appendix}
\renewcommand{\thefigure}{A\arabic{figure}}
\setcounter{figure}{0}
\begin{figure}[h]
	\begin{center}
	\includegraphics[width=0.95\columnwidth]{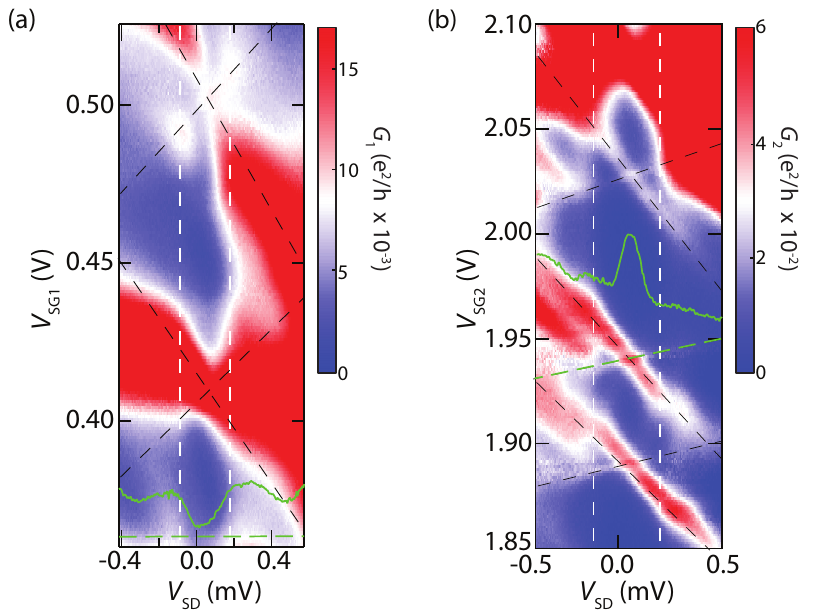}
    \end{center} 	
	\caption{(a) Differential conductance $G_1$ of $\textrm{QD}_1$ as a function of $\textrm{SG}_1$ and source drain bias $\textrm{V}_{SD}$. White dashed lines indicate a suppression of conductance due to the superconducting energy gap. Cross section along green dashed line shown in green. (b)  Differential conductance $G_2$ of $\textrm{QD}_2$ as a function of $\textrm{SG}_2$ and $\textrm{V}_{SD}$ respectively. Cross section along green dashed line shown in green, representing an enhancement of Andreev Reflection due to the stronger coupling of $\textrm{QD}_2$ to the SC.}
	\label{fig:D_diamonds}
\end{figure}
\begin{figure}[h]
	\begin{center}
	\includegraphics[width=0.95\columnwidth]{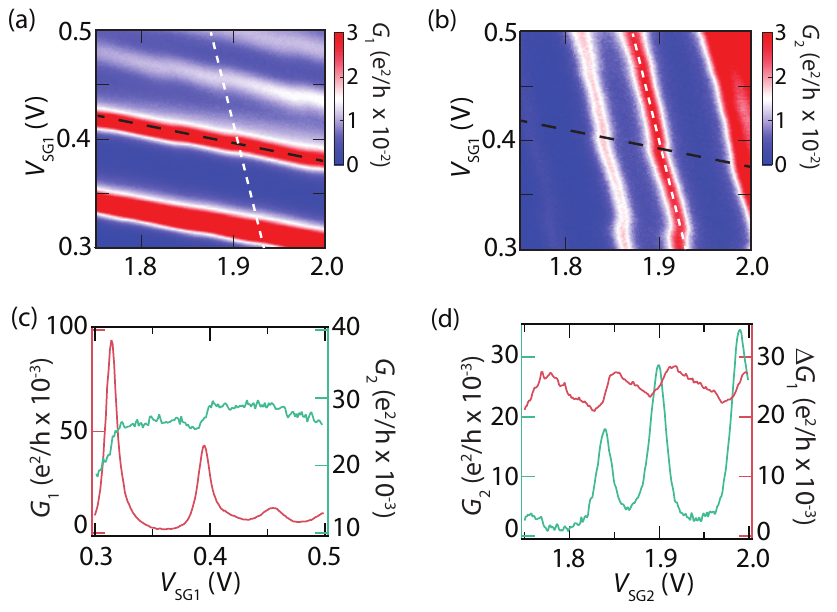}
    \end{center} 	
	\caption{(a,b) Differential conductance $G_1$ and $G_2$ of $\textrm{QD}_1$ and $\textrm{QD}_2$ as a function of $V_{SG1}$ and $V_{SG2}$. During this measurements a constant out-of-plane external magnetic field of \SI{250}{\milli \tesla} was applied. (c,d) Cross sections along the dashed white and black lines shown in (a) and (b). The positive correlation, which was evident in Fig.\ref{fig:CPS-gate} in the main text, is now either absent or replaced by a negative correlation. The latter is expected for classical correlations that can be described by a simple resistor network.\cite{Hofstetter2009} In (d) a linear background was removed from $G_1$, therefore its denoted as $\Delta G_1$. }
	\label{fig:E_CPS_Bfield}
\end{figure}
\newpage
\bibliography{literature}

\providecommand{\noopsort}[1]{}\providecommand{\singleletter}[1]{#1}%
\begin{thebibliography}{36}%
\makeatletter
\providecommand \@ifxundefined [1]{%
 \@ifx{#1\undefined}
}%
\providecommand \@ifnum [1]{%
 \ifnum #1\expandafter \@firstoftwo
 \else \expandafter \@secondoftwo
 \fi
}%
\providecommand \@ifx [1]{%
 \ifx #1\expandafter \@firstoftwo
 \else \expandafter \@secondoftwo
 \fi
}%
\providecommand \natexlab [1]{#1}%
\providecommand \enquote  [1]{``#1''}%
\providecommand \bibnamefont  [1]{#1}%
\providecommand \bibfnamefont [1]{#1}%
\providecommand \citenamefont [1]{#1}%
\providecommand \href@noop [0]{\@secondoftwo}%
\providecommand \href [0]{\begingroup \@sanitize@url \@href}%
\providecommand \@href[1]{\@@startlink{#1}\@@href}%
\providecommand \@@href[1]{\endgroup#1\@@endlink}%
\providecommand \@sanitize@url [0]{\catcode `\\12\catcode `\$12\catcode
  `\&12\catcode `\#12\catcode `\^12\catcode `\_12\catcode `\%12\relax}%
\providecommand \@@startlink[1]{}%
\providecommand \@@endlink[0]{}%
\providecommand \url  [0]{\begingroup\@sanitize@url \@url }%
\providecommand \@url [1]{\endgroup\@href {#1}{\urlprefix }}%
\providecommand \urlprefix  [0]{URL }%
\providecommand \Eprint [0]{\href }%
\providecommand \doibase [0]{http://dx.doi.org/}%
\providecommand \selectlanguage [0]{\@gobble}%
\providecommand \bibinfo  [0]{\@secondoftwo}%
\providecommand \bibfield  [0]{\@secondoftwo}%
\providecommand \translation [1]{[#1]}%
\providecommand \BibitemOpen [0]{}%
\providecommand \bibitemStop [0]{}%
\providecommand \bibitemNoStop [0]{.\EOS\space}%
\providecommand \EOS [0]{\spacefactor3000\relax}%
\providecommand \BibitemShut  [1]{\csname bibitem#1\endcsname}%
\let\auto@bib@innerbib\@empty
\bibitem [{\citenamefont {Sarma}, \citenamefont {Freedman},\ and\ \citenamefont
  {Nayak}(2015)}]{Sarma2015}%
  \BibitemOpen
  \bibfield  {author} {\bibinfo {author} {\bibfnamefont {S.~D.}\ \bibnamefont
  {Sarma}}, \bibinfo {author} {\bibfnamefont {M.}~\bibnamefont {Freedman}}, \
  and\ \bibinfo {author} {\bibfnamefont {C.}~\bibnamefont {Nayak}},\ }\bibfield
   {title} {\enquote {\bibinfo {title} {Majorana zero modes and topological
  quantum computation},}\ }\href {\doibase 10.1038/npjqi.2015.1} {\bibfield
  {journal} {\bibinfo  {journal} {npj Quantum Information}\ }\textbf {\bibinfo
  {volume} {1}},\ \bibinfo {pages} {15001} (\bibinfo {year}
  {2015})}\BibitemShut {NoStop}%
\bibitem [{\citenamefont {Hoffman}\ \emph {et~al.}(2016)\citenamefont
  {Hoffman}, \citenamefont {Schrade}, \citenamefont {Klinovaja},\ and\
  \citenamefont {Loss}}]{Hoffman2016}%
  \BibitemOpen
  \bibfield  {author} {\bibinfo {author} {\bibfnamefont {S.}~\bibnamefont
  {Hoffman}}, \bibinfo {author} {\bibfnamefont {C.}~\bibnamefont {Schrade}},
  \bibinfo {author} {\bibfnamefont {J.}~\bibnamefont {Klinovaja}}, \ and\
  \bibinfo {author} {\bibfnamefont {D.}~\bibnamefont {Loss}},\ }\bibfield
  {title} {\enquote {\bibinfo {title} {Universal quantum computation with
  hybrid spin-majorana qubits},}\ }\href {\doibase 10.1103/physrevb.94.045316}
  {\bibfield  {journal} {\bibinfo  {journal} {Physical Review B}\ }\textbf
  {\bibinfo {volume} {94}},\ \bibinfo {pages} {045316} (\bibinfo {year}
  {2016})}\BibitemShut {NoStop}%
\bibitem [{\citenamefont {Mourik}\ \emph {et~al.}(2012)\citenamefont {Mourik},
  \citenamefont {Zuo}, \citenamefont {Frolov}, \citenamefont {Plissard},
  \citenamefont {Bakkers},\ and\ \citenamefont {Kouwenhoven}}]{Mourik2012}%
  \BibitemOpen
  \bibfield  {author} {\bibinfo {author} {\bibfnamefont {V.}~\bibnamefont
  {Mourik}}, \bibinfo {author} {\bibfnamefont {K.}~\bibnamefont {Zuo}},
  \bibinfo {author} {\bibfnamefont {S.~M.}\ \bibnamefont {Frolov}}, \bibinfo
  {author} {\bibfnamefont {S.~R.}\ \bibnamefont {Plissard}}, \bibinfo {author}
  {\bibfnamefont {E.~P. A.~M.}\ \bibnamefont {Bakkers}}, \ and\ \bibinfo
  {author} {\bibfnamefont {L.~P.}\ \bibnamefont {Kouwenhoven}},\ }\bibfield
  {title} {\enquote {\bibinfo {title} {Signatures of majorana fermions in
  hybrid superconductor-semiconductor nanowire devices},}\ }\href {\doibase
  10.1126/science.1222360} {\bibfield  {journal} {\bibinfo  {journal}
  {Science}\ }\textbf {\bibinfo {volume} {336}},\ \bibinfo {pages} {1003--1007}
  (\bibinfo {year} {2012})}\BibitemShut {NoStop}%
\bibitem [{\citenamefont {Deng}\ \emph {et~al.}(2012)\citenamefont {Deng},
  \citenamefont {Yu}, \citenamefont {Huang}, \citenamefont {Larsson},
  \citenamefont {Caroff},\ and\ \citenamefont {Xu}}]{Deng2012}%
  \BibitemOpen
  \bibfield  {author} {\bibinfo {author} {\bibfnamefont {M.~T.}\ \bibnamefont
  {Deng}}, \bibinfo {author} {\bibfnamefont {C.~L.}\ \bibnamefont {Yu}},
  \bibinfo {author} {\bibfnamefont {G.~Y.}\ \bibnamefont {Huang}}, \bibinfo
  {author} {\bibfnamefont {M.}~\bibnamefont {Larsson}}, \bibinfo {author}
  {\bibfnamefont {P.}~\bibnamefont {Caroff}}, \ and\ \bibinfo {author}
  {\bibfnamefont {H.~Q.}\ \bibnamefont {Xu}},\ }\bibfield  {title} {\enquote
  {\bibinfo {title} {Anomalous zero-bias conductance peak in a
  nb{\textendash}{InSb} nanowire{\textendash}nb hybrid device},}\ }\href
  {\doibase 10.1021/nl303758w} {\bibfield  {journal} {\bibinfo  {journal} {Nano
  Letters}\ }\textbf {\bibinfo {volume} {12}},\ \bibinfo {pages} {6414--6419}
  (\bibinfo {year} {2012})}\BibitemShut {NoStop}%
\bibitem [{\citenamefont {Albrecht}\ \emph {et~al.}(2016)\citenamefont
  {Albrecht}, \citenamefont {Higginbotham}, \citenamefont {Madsen},
  \citenamefont {Kuemmeth}, \citenamefont {Jespersen}, \citenamefont
  {Nyg{\aa}rd}, \citenamefont {Krogstrup},\ and\ \citenamefont
  {Marcus}}]{Albrecht2016}%
  \BibitemOpen
  \bibfield  {author} {\bibinfo {author} {\bibfnamefont {S.~M.}\ \bibnamefont
  {Albrecht}}, \bibinfo {author} {\bibfnamefont {A.~P.}\ \bibnamefont
  {Higginbotham}}, \bibinfo {author} {\bibfnamefont {M.}~\bibnamefont
  {Madsen}}, \bibinfo {author} {\bibfnamefont {F.}~\bibnamefont {Kuemmeth}},
  \bibinfo {author} {\bibfnamefont {T.~S.}\ \bibnamefont {Jespersen}}, \bibinfo
  {author} {\bibfnamefont {J.}~\bibnamefont {Nyg{\aa}rd}}, \bibinfo {author}
  {\bibfnamefont {P.}~\bibnamefont {Krogstrup}}, \ and\ \bibinfo {author}
  {\bibfnamefont {C.~M.}\ \bibnamefont {Marcus}},\ }\bibfield  {title}
  {\enquote {\bibinfo {title} {Exponential protection of zero modes in majorana
  islands},}\ }\href {\doibase 10.1038/nature17162} {\bibfield  {journal}
  {\bibinfo  {journal} {Nature}\ }\textbf {\bibinfo {volume} {531}},\ \bibinfo
  {pages} {206--209} (\bibinfo {year} {2016})}\BibitemShut {NoStop}%
\bibitem [{\citenamefont {Deng}\ \emph {et~al.}(2016)\citenamefont {Deng},
  \citenamefont {Vaitiek{\.{e}}nas}, \citenamefont {Hansen}, \citenamefont
  {Danon}, \citenamefont {Leijnse}, \citenamefont {Flensberg}, \citenamefont
  {Nyg{\aa}rd}, \citenamefont {Krogstrup},\ and\ \citenamefont
  {Marcus}}]{Deng2016}%
  \BibitemOpen
  \bibfield  {author} {\bibinfo {author} {\bibfnamefont {M.~T.}\ \bibnamefont
  {Deng}}, \bibinfo {author} {\bibfnamefont {S.}~\bibnamefont
  {Vaitiek{\.{e}}nas}}, \bibinfo {author} {\bibfnamefont {E.~B.}\ \bibnamefont
  {Hansen}}, \bibinfo {author} {\bibfnamefont {J.}~\bibnamefont {Danon}},
  \bibinfo {author} {\bibfnamefont {M.}~\bibnamefont {Leijnse}}, \bibinfo
  {author} {\bibfnamefont {K.}~\bibnamefont {Flensberg}}, \bibinfo {author}
  {\bibfnamefont {J.}~\bibnamefont {Nyg{\aa}rd}}, \bibinfo {author}
  {\bibfnamefont {P.}~\bibnamefont {Krogstrup}}, \ and\ \bibinfo {author}
  {\bibfnamefont {C.~M.}\ \bibnamefont {Marcus}},\ }\bibfield  {title}
  {\enquote {\bibinfo {title} {Majorana bound state in a coupled quantum-dot
  hybrid-nanowire system},}\ }\href {\doibase 10.1126/science.aaf3961}
  {\bibfield  {journal} {\bibinfo  {journal} {Science}\ }\textbf {\bibinfo
  {volume} {354}},\ \bibinfo {pages} {1557--1562} (\bibinfo {year}
  {2016})}\BibitemShut {NoStop}%
\bibitem [{\citenamefont {G{\"u}l}\ \emph {et~al.}(2018)\citenamefont
  {G{\"u}l}, \citenamefont {Zhang}, \citenamefont {Bommer}, \citenamefont
  {de~Moor}, \citenamefont {Car}, \citenamefont {Plissard}, \citenamefont
  {Bakkers}, \citenamefont {Geresdi}, \citenamefont {Watanabe}, \citenamefont
  {Taniguchi},\ and\ \citenamefont {Kouwenhoven}}]{Guel2018}%
  \BibitemOpen
  \bibfield  {author} {\bibinfo {author} {\bibfnamefont {{\"O}.}~\bibnamefont
  {G{\"u}l}}, \bibinfo {author} {\bibfnamefont {H.}~\bibnamefont {Zhang}},
  \bibinfo {author} {\bibfnamefont {J.~D.~S.}\ \bibnamefont {Bommer}}, \bibinfo
  {author} {\bibfnamefont {M.~W.~A.}\ \bibnamefont {de~Moor}}, \bibinfo
  {author} {\bibfnamefont {D.}~\bibnamefont {Car}}, \bibinfo {author}
  {\bibfnamefont {S.~R.}\ \bibnamefont {Plissard}}, \bibinfo {author}
  {\bibfnamefont {E.~P. A.~M.}\ \bibnamefont {Bakkers}}, \bibinfo {author}
  {\bibfnamefont {A.}~\bibnamefont {Geresdi}}, \bibinfo {author} {\bibfnamefont
  {K.}~\bibnamefont {Watanabe}}, \bibinfo {author} {\bibfnamefont
  {T.}~\bibnamefont {Taniguchi}}, \ and\ \bibinfo {author} {\bibfnamefont
  {L.~P.}\ \bibnamefont {Kouwenhoven}},\ }\bibfield  {title} {\enquote
  {\bibinfo {title} {Ballistic majorana nanowire devices},}\ }\href {\doibase
  10.1038/s41565-017-0032-8} {\bibfield  {journal} {\bibinfo  {journal} {Nature
  Nanotechnology}\ } (\bibinfo {year} {2018}),\
  10.1038/s41565-017-0032-8}\BibitemShut {NoStop}%
\bibitem [{\citenamefont {Alicea}(2010)}]{Alicea2010}%
  \BibitemOpen
  \bibfield  {author} {\bibinfo {author} {\bibfnamefont {J.}~\bibnamefont
  {Alicea}},\ }\bibfield  {title} {\enquote {\bibinfo {title} {Majorana
  fermions in a tunable semiconductor device},}\ }\href {\doibase
  10.1103/physrevb.81.125318} {\bibfield  {journal} {\bibinfo  {journal}
  {Physical Review B}\ }\textbf {\bibinfo {volume} {81}},\ \bibinfo {pages}
  {125318} (\bibinfo {year} {2010})}\BibitemShut {NoStop}%
\bibitem [{\citenamefont {Oreg}, \citenamefont {Refael},\ and\ \citenamefont
  {von Oppen}(2010)}]{Oreg2010}%
  \BibitemOpen
  \bibfield  {author} {\bibinfo {author} {\bibfnamefont {Y.}~\bibnamefont
  {Oreg}}, \bibinfo {author} {\bibfnamefont {G.}~\bibnamefont {Refael}}, \ and\
  \bibinfo {author} {\bibfnamefont {F.}~\bibnamefont {von Oppen}},\ }\bibfield
  {title} {\enquote {\bibinfo {title} {Helical liquids and majorana bound
  states in quantum wires},}\ }\href {\doibase 10.1103/physrevlett.105.177002}
  {\bibfield  {journal} {\bibinfo  {journal} {Physical Review Letters}\
  }\textbf {\bibinfo {volume} {105}},\ \bibinfo {pages} {177002} (\bibinfo
  {year} {2010})}\BibitemShut {NoStop}%
\bibitem [{\citenamefont {Pachos}(2012)}]{Pachos2012}%
  \BibitemOpen
  \bibfield  {author} {\bibinfo {author} {\bibfnamefont {J.~K.}\ \bibnamefont
  {Pachos}},\ }\href@noop {} {\emph {\bibinfo {title} {Introduction to
  Topological Quantum Computation}}}\ (\bibinfo  {publisher} {Cambridge
  University Press},\ \bibinfo {year} {2012})\BibitemShut {NoStop}%
\bibitem [{\citenamefont {Keselman}\ \emph {et~al.}(2013)\citenamefont
  {Keselman}, \citenamefont {Fu}, \citenamefont {Stern},\ and\ \citenamefont
  {Berg}}]{Keselman2013}%
  \BibitemOpen
  \bibfield  {author} {\bibinfo {author} {\bibfnamefont {A.}~\bibnamefont
  {Keselman}}, \bibinfo {author} {\bibfnamefont {L.}~\bibnamefont {Fu}},
  \bibinfo {author} {\bibfnamefont {A.}~\bibnamefont {Stern}}, \ and\ \bibinfo
  {author} {\bibfnamefont {E.}~\bibnamefont {Berg}},\ }\bibfield  {title}
  {\enquote {\bibinfo {title} {Inducing time-reversal-invariant topological
  superconductivity and fermion parity pumping in quantum wires},}\ }\href
  {\doibase 10.1103/physrevlett.111.116402} {\bibfield  {journal} {\bibinfo
  {journal} {Physical Review Letters}\ }\textbf {\bibinfo {volume} {111}},\
  \bibinfo {pages} {116402} (\bibinfo {year} {2013})}\BibitemShut {NoStop}%
\bibitem [{\citenamefont {Klinovaja}\ and\ \citenamefont
  {Loss}(2014)}]{Klinovaja2014}%
  \BibitemOpen
  \bibfield  {author} {\bibinfo {author} {\bibfnamefont {J.}~\bibnamefont
  {Klinovaja}}\ and\ \bibinfo {author} {\bibfnamefont {D.}~\bibnamefont
  {Loss}},\ }\bibfield  {title} {\enquote {\bibinfo {title} {Time-reversal
  invariant parafermions in interacting rashba nanowires},}\ }\href {\doibase
  10.1103/physrevb.90.045118} {\bibfield  {journal} {\bibinfo  {journal}
  {Physical Review B}\ }\textbf {\bibinfo {volume} {90}},\ \bibinfo {pages}
  {045118} (\bibinfo {year} {2014})}\BibitemShut {NoStop}%
\bibitem [{\citenamefont {Gaidamauskas}, \citenamefont {Paaske},\ and\
  \citenamefont {Flensberg}(2014)}]{Gaidamauskas2014}%
  \BibitemOpen
  \bibfield  {author} {\bibinfo {author} {\bibfnamefont {E.}~\bibnamefont
  {Gaidamauskas}}, \bibinfo {author} {\bibfnamefont {J.}~\bibnamefont
  {Paaske}}, \ and\ \bibinfo {author} {\bibfnamefont {K.}~\bibnamefont
  {Flensberg}},\ }\bibfield  {title} {\enquote {\bibinfo {title} {Majorana
  bound states in two-channel time-reversal-symmetric nanowire systems},}\
  }\href {\doibase 10.1103/PhysRevLett.112.126402} {\bibfield  {journal}
  {\bibinfo  {journal} {Physical Review Letters}\ }\textbf {\bibinfo {volume}
  {112}},\ \bibinfo {pages} {233513} (\bibinfo {year} {2014})}\BibitemShut
  {NoStop}%
\bibitem [{\citenamefont {Sato}, \citenamefont {Loss},\ and\ \citenamefont
  {Tserkovnyak}(2012)}]{Sato2012}%
  \BibitemOpen
  \bibfield  {author} {\bibinfo {author} {\bibfnamefont {K.}~\bibnamefont
  {Sato}}, \bibinfo {author} {\bibfnamefont {D.}~\bibnamefont {Loss}}, \ and\
  \bibinfo {author} {\bibfnamefont {Y.}~\bibnamefont {Tserkovnyak}},\
  }\bibfield  {title} {\enquote {\bibinfo {title} {Crossed andreev reflection
  in quantum wires with strong spin-orbit interaction},}\ }\href {\doibase
  10.1103/physrevb.85.235433} {\bibfield  {journal} {\bibinfo  {journal}
  {Physical Review B}\ }\textbf {\bibinfo {volume} {85}},\ \bibinfo {pages}
  {235433} (\bibinfo {year} {2012})}\BibitemShut {NoStop}%
\bibitem [{\citenamefont {Bena}\ \emph {et~al.}(2002)\citenamefont {Bena},
  \citenamefont {Vishveshwara}, \citenamefont {Balents},\ and\ \citenamefont
  {Fisher}}]{Bena2002}%
  \BibitemOpen
  \bibfield  {author} {\bibinfo {author} {\bibfnamefont {C.}~\bibnamefont
  {Bena}}, \bibinfo {author} {\bibfnamefont {S.}~\bibnamefont {Vishveshwara}},
  \bibinfo {author} {\bibfnamefont {L.}~\bibnamefont {Balents}}, \ and\
  \bibinfo {author} {\bibfnamefont {M.~P.~A.}\ \bibnamefont {Fisher}},\
  }\bibfield  {title} {\enquote {\bibinfo {title} {Quantum entanglement in
  carbon nanotubes},}\ }\href {\doibase 10.1103/physrevlett.89.037901}
  {\bibfield  {journal} {\bibinfo  {journal} {Physical Review Letters}\
  }\textbf {\bibinfo {volume} {89}},\ \bibinfo {pages} {037901} (\bibinfo
  {year} {2002})}\BibitemShut {NoStop}%
\bibitem [{\citenamefont {Recher}\ and\ \citenamefont
  {Loss}(2002)}]{Recher2002}%
  \BibitemOpen
  \bibfield  {author} {\bibinfo {author} {\bibfnamefont {P.}~\bibnamefont
  {Recher}}\ and\ \bibinfo {author} {\bibfnamefont {D.}~\bibnamefont {Loss}},\
  }\bibfield  {title} {\enquote {\bibinfo {title} {Superconductor coupled to
  two luttinger liquids as an entangler for electron spins},}\ }\href {\doibase
  10.1103/physrevb.65.165327} {\bibfield  {journal} {\bibinfo  {journal}
  {Physical Review B}\ }\textbf {\bibinfo {volume} {65}},\ \bibinfo {pages}
  {165327} (\bibinfo {year} {2002})}\BibitemShut {NoStop}%
\bibitem [{\citenamefont {Thakurathi}\ \emph {et~al.}(2018)\citenamefont
  {Thakurathi}, \citenamefont {Simon}, \citenamefont {Mandal}, \citenamefont
  {Klinovaja},\ and\ \citenamefont {Loss}}]{Thakurathi2018}%
  \BibitemOpen
  \bibfield  {author} {\bibinfo {author} {\bibfnamefont {M.}~\bibnamefont
  {Thakurathi}}, \bibinfo {author} {\bibfnamefont {P.}~\bibnamefont {Simon}},
  \bibinfo {author} {\bibfnamefont {I.}~\bibnamefont {Mandal}}, \bibinfo
  {author} {\bibfnamefont {J.}~\bibnamefont {Klinovaja}}, \ and\ \bibinfo
  {author} {\bibfnamefont {D.}~\bibnamefont {Loss}},\ }\bibfield  {title}
  {\enquote {\bibinfo {title} {Majorana kramers pairs in rashba double
  nanowires with interactions and disorder},}\ }\href {\doibase
  10.1103/physrevb.97.045415} {\bibfield  {journal} {\bibinfo  {journal}
  {Physical Review B}\ }\textbf {\bibinfo {volume} {97}},\ \bibinfo {pages}
  {045415} (\bibinfo {year} {2018})}\BibitemShut {NoStop}%
\bibitem [{\citenamefont {Recher}, \citenamefont {Sukhorukov},\ and\
  \citenamefont {Loss}(2001)}]{Recher2001}%
  \BibitemOpen
  \bibfield  {author} {\bibinfo {author} {\bibfnamefont {P.}~\bibnamefont
  {Recher}}, \bibinfo {author} {\bibfnamefont {E.~V.}\ \bibnamefont
  {Sukhorukov}}, \ and\ \bibinfo {author} {\bibfnamefont {D.}~\bibnamefont
  {Loss}},\ }\bibfield  {title} {\enquote {\bibinfo {title} {Andreev tunneling,
  coulomb blockade, and resonant transport of nonlocal spin-entangled
  electrons},}\ }\href {\doibase 10.1103/physrevb.63.165314} {\bibfield
  {journal} {\bibinfo  {journal} {Physical Review B}\ }\textbf {\bibinfo
  {volume} {63}},\ \bibinfo {pages} {165314} (\bibinfo {year}
  {2001})}\BibitemShut {NoStop}%
\bibitem [{\citenamefont {Chevallier}\ \emph {et~al.}(2011)\citenamefont
  {Chevallier}, \citenamefont {Rech}, \citenamefont {Jonckheere},\ and\
  \citenamefont {Martin}}]{Chevallier2012}%
  \BibitemOpen
  \bibfield  {author} {\bibinfo {author} {\bibfnamefont {D.}~\bibnamefont
  {Chevallier}}, \bibinfo {author} {\bibfnamefont {J.}~\bibnamefont {Rech}},
  \bibinfo {author} {\bibfnamefont {T.}~\bibnamefont {Jonckheere}}, \ and\
  \bibinfo {author} {\bibfnamefont {T.}~\bibnamefont {Martin}},\ }\bibfield
  {title} {\enquote {\bibinfo {title} {Current and noise correlations in a
  double-dot cooper-pair beam splitter},}\ }\href {\doibase
  10.1103/PhysRevB.83.125421} {\bibfield  {journal} {\bibinfo  {journal} {Phys.
  Rev. B}\ }\textbf {\bibinfo {volume} {83}},\ \bibinfo {pages} {125421}
  (\bibinfo {year} {2011})}\BibitemShut {NoStop}%
\bibitem [{\citenamefont {Hofstetter}\ \emph {et~al.}(2009)\citenamefont
  {Hofstetter}, \citenamefont {Csonka}, \citenamefont {Nyg{\aa}rd},\ and\
  \citenamefont {Sch{\"o}nenberger}}]{Hofstetter2009}%
  \BibitemOpen
  \bibfield  {author} {\bibinfo {author} {\bibfnamefont {L.}~\bibnamefont
  {Hofstetter}}, \bibinfo {author} {\bibfnamefont {S.}~\bibnamefont {Csonka}},
  \bibinfo {author} {\bibfnamefont {J.}~\bibnamefont {Nyg{\aa}rd}}, \ and\
  \bibinfo {author} {\bibfnamefont {C.}~\bibnamefont {Sch{\"o}nenberger}},\
  }\bibfield  {title} {\enquote {\bibinfo {title} {Cooper pair splitter
  realized in a two-quantum-dot y-junction},}\ }\href {\doibase
  10.1038/nature08432} {\bibfield  {journal} {\bibinfo  {journal} {Nature}\
  }\textbf {\bibinfo {volume} {461}},\ \bibinfo {pages} {960--963} (\bibinfo
  {year} {2009})}\BibitemShut {NoStop}%
\bibitem [{\citenamefont {Hofstetter}\ \emph {et~al.}(2011)\citenamefont
  {Hofstetter}, \citenamefont {Csonka}, \citenamefont {Baumgartner},
  \citenamefont {Fülöp}, \citenamefont {d'Hollosy}, \citenamefont
  {Nyg{\aa}rd},\ and\ \citenamefont {Sch{\"o}nenberger}}]{Hofstetter2011}%
  \BibitemOpen
  \bibfield  {author} {\bibinfo {author} {\bibfnamefont {L.}~\bibnamefont
  {Hofstetter}}, \bibinfo {author} {\bibfnamefont {S.}~\bibnamefont {Csonka}},
  \bibinfo {author} {\bibfnamefont {A.}~\bibnamefont {Baumgartner}}, \bibinfo
  {author} {\bibfnamefont {G.}~\bibnamefont {Fülöp}}, \bibinfo {author}
  {\bibfnamefont {S.}~\bibnamefont {d'Hollosy}}, \bibinfo {author}
  {\bibfnamefont {J.}~\bibnamefont {Nyg{\aa}rd}}, \ and\ \bibinfo {author}
  {\bibfnamefont {C.}~\bibnamefont {Sch{\"o}nenberger}},\ }\bibfield  {title}
  {\enquote {\bibinfo {title} {Finite-bias cooper pair splitting},}\ }\href
  {\doibase 10.1103/physrevlett.107.136801} {\bibfield  {journal} {\bibinfo
  {journal} {Physical Review Letters}\ }\textbf {\bibinfo {volume} {107}},\
  \bibinfo {pages} {136801} (\bibinfo {year} {2011})}\BibitemShut {NoStop}%
\bibitem [{\citenamefont {Das}\ \emph {et~al.}(2012)\citenamefont {Das},
  \citenamefont {Ronen}, \citenamefont {Heiblum}, \citenamefont {Mahalu},
  \citenamefont {Kretinin},\ and\ \citenamefont {Shtrikman}}]{Das2012}%
  \BibitemOpen
  \bibfield  {author} {\bibinfo {author} {\bibfnamefont {A.}~\bibnamefont
  {Das}}, \bibinfo {author} {\bibfnamefont {Y.}~\bibnamefont {Ronen}}, \bibinfo
  {author} {\bibfnamefont {M.}~\bibnamefont {Heiblum}}, \bibinfo {author}
  {\bibfnamefont {D.}~\bibnamefont {Mahalu}}, \bibinfo {author} {\bibfnamefont
  {A.~V.}\ \bibnamefont {Kretinin}}, \ and\ \bibinfo {author} {\bibfnamefont
  {H.}~\bibnamefont {Shtrikman}},\ }\bibfield  {title} {\enquote {\bibinfo
  {title} {High-efficiency cooper pair splitting demonstrated by two-particle
  conductance resonance and positive noise cross-correlation},}\ }\href
  {\doibase 10.1038/ncomms2169} {\bibfield  {journal} {\bibinfo  {journal}
  {Nature Communications}\ }\textbf {\bibinfo {volume} {3}},\ \bibinfo {pages}
  {2169} (\bibinfo {year} {2012})}\BibitemShut {NoStop}%
\bibitem [{\citenamefont {F{\"u}l{\"o}p}\ \emph {et~al.}(2014)\citenamefont
  {F{\"u}l{\"o}p}, \citenamefont {d'Hollosy}, \citenamefont {Baumgartner},
  \citenamefont {Makk}, \citenamefont {Guzenko}, \citenamefont {Madsen},
  \citenamefont {Nyg{\aa}rd}, \citenamefont {Sch{\"o}nenberger},\ and\
  \citenamefont {Csonka}}]{Fueloep2014}%
  \BibitemOpen
  \bibfield  {author} {\bibinfo {author} {\bibfnamefont {G.}~\bibnamefont
  {F{\"u}l{\"o}p}}, \bibinfo {author} {\bibfnamefont {S.}~\bibnamefont
  {d'Hollosy}}, \bibinfo {author} {\bibfnamefont {A.}~\bibnamefont
  {Baumgartner}}, \bibinfo {author} {\bibfnamefont {P.}~\bibnamefont {Makk}},
  \bibinfo {author} {\bibfnamefont {V.~A.}\ \bibnamefont {Guzenko}}, \bibinfo
  {author} {\bibfnamefont {M.~H.}\ \bibnamefont {Madsen}}, \bibinfo {author}
  {\bibfnamefont {J.}~\bibnamefont {Nyg{\aa}rd}}, \bibinfo {author}
  {\bibfnamefont {C.}~\bibnamefont {Sch{\"o}nenberger}}, \ and\ \bibinfo
  {author} {\bibfnamefont {S.}~\bibnamefont {Csonka}},\ }\bibfield  {title}
  {\enquote {\bibinfo {title} {Local electrical tuning of the nonlocal signals
  in a cooper pair splitter},}\ }\href {\doibase 10.1103/physrevb.90.235412}
  {\bibfield  {journal} {\bibinfo  {journal} {Physical Review B}\ }\textbf
  {\bibinfo {volume} {90}},\ \bibinfo {pages} {235412} (\bibinfo {year}
  {2014})}\BibitemShut {NoStop}%
\bibitem [{\citenamefont {F{\"u}l{\"o}p}\ \emph {et~al.}(2015)\citenamefont
  {F{\"u}l{\"o}p}, \citenamefont {Dom{\'{\i}}nguez}, \citenamefont {d'Hollosy},
  \citenamefont {Baumgartner}, \citenamefont {Makk}, \citenamefont {Madsen},
  \citenamefont {Guzenko}, \citenamefont {Nyg{\aa}rd}, \citenamefont
  {Sch{\"o}nenberger}, \citenamefont {Yeyati},\ and\ \citenamefont
  {Csonka}}]{Fueloep2015}%
  \BibitemOpen
  \bibfield  {author} {\bibinfo {author} {\bibfnamefont {G.}~\bibnamefont
  {F{\"u}l{\"o}p}}, \bibinfo {author} {\bibfnamefont {F.}~\bibnamefont
  {Dom{\'{\i}}nguez}}, \bibinfo {author} {\bibfnamefont {S.}~\bibnamefont
  {d'Hollosy}}, \bibinfo {author} {\bibfnamefont {A.}~\bibnamefont
  {Baumgartner}}, \bibinfo {author} {\bibfnamefont {P.}~\bibnamefont {Makk}},
  \bibinfo {author} {\bibfnamefont {M.}~\bibnamefont {Madsen}}, \bibinfo
  {author} {\bibfnamefont {V.}~\bibnamefont {Guzenko}}, \bibinfo {author}
  {\bibfnamefont {J.}~\bibnamefont {Nyg{\aa}rd}}, \bibinfo {author}
  {\bibfnamefont {C.}~\bibnamefont {Sch{\"o}nenberger}}, \bibinfo {author}
  {\bibfnamefont {A.~L.}\ \bibnamefont {Yeyati}}, \ and\ \bibinfo {author}
  {\bibfnamefont {S.}~\bibnamefont {Csonka}},\ }\bibfield  {title} {\enquote
  {\bibinfo {title} {Magnetic field tuning and quantum interference in a cooper
  pair splitter},}\ }\href {\doibase 10.1103/physrevlett.115.227003} {\bibfield
   {journal} {\bibinfo  {journal} {Physical Review Letters}\ }\textbf {\bibinfo
  {volume} {115}},\ \bibinfo {pages} {227003} (\bibinfo {year}
  {2015})}\BibitemShut {NoStop}%
\bibitem [{\citenamefont {Herrmann}\ \emph {et~al.}(2010)\citenamefont
  {Herrmann}, \citenamefont {Portier}, \citenamefont {Roche}, \citenamefont
  {Yeyati}, \citenamefont {Kontos},\ and\ \citenamefont
  {Strunk}}]{Herrmann2010}%
  \BibitemOpen
  \bibfield  {author} {\bibinfo {author} {\bibfnamefont {L.~G.}\ \bibnamefont
  {Herrmann}}, \bibinfo {author} {\bibfnamefont {F.}~\bibnamefont {Portier}},
  \bibinfo {author} {\bibfnamefont {P.}~\bibnamefont {Roche}}, \bibinfo
  {author} {\bibfnamefont {A.~L.}\ \bibnamefont {Yeyati}}, \bibinfo {author}
  {\bibfnamefont {T.}~\bibnamefont {Kontos}}, \ and\ \bibinfo {author}
  {\bibfnamefont {C.}~\bibnamefont {Strunk}},\ }\bibfield  {title} {\enquote
  {\bibinfo {title} {Carbon nanotubes as cooper-pair beam splitters},}\ }\href
  {\doibase 10.1103/physrevlett.104.026801} {\bibfield  {journal} {\bibinfo
  {journal} {Physical Review Letters}\ }\textbf {\bibinfo {volume} {104}},\
  \bibinfo {pages} {026801} (\bibinfo {year} {2010})}\BibitemShut {NoStop}%
\bibitem [{\citenamefont {Schindele}, \citenamefont {Baumgartner},\ and\
  \citenamefont {Sch{\"o}nenberger}(2012)}]{Schindele2012}%
  \BibitemOpen
  \bibfield  {author} {\bibinfo {author} {\bibfnamefont {J.}~\bibnamefont
  {Schindele}}, \bibinfo {author} {\bibfnamefont {A.}~\bibnamefont
  {Baumgartner}}, \ and\ \bibinfo {author} {\bibfnamefont {C.}~\bibnamefont
  {Sch{\"o}nenberger}},\ }\bibfield  {title} {\enquote {\bibinfo {title}
  {Near-unity cooper pair splitting efficiency},}\ }\href {\doibase
  10.1103/physrevlett.109.157002} {\bibfield  {journal} {\bibinfo  {journal}
  {Physical Review Letters}\ }\textbf {\bibinfo {volume} {109}},\ \bibinfo
  {pages} {157002} (\bibinfo {year} {2012})}\BibitemShut {NoStop}%
\bibitem [{\citenamefont {Tan}\ \emph {et~al.}(2015)\citenamefont {Tan},
  \citenamefont {Cox}, \citenamefont {Nieminen}, \citenamefont {Lähteenmäki},
  \citenamefont {Golubev}, \citenamefont {Lesovik},\ and\ \citenamefont
  {Hakonen}}]{Tan2015}%
  \BibitemOpen
  \bibfield  {author} {\bibinfo {author} {\bibfnamefont {Z.}~\bibnamefont
  {Tan}}, \bibinfo {author} {\bibfnamefont {D.}~\bibnamefont {Cox}}, \bibinfo
  {author} {\bibfnamefont {T.}~\bibnamefont {Nieminen}}, \bibinfo {author}
  {\bibfnamefont {P.}~\bibnamefont {Lähteenmäki}}, \bibinfo {author}
  {\bibfnamefont {D.}~\bibnamefont {Golubev}}, \bibinfo {author} {\bibfnamefont
  {G.}~\bibnamefont {Lesovik}}, \ and\ \bibinfo {author} {\bibfnamefont
  {P.}~\bibnamefont {Hakonen}},\ }\bibfield  {title} {\enquote {\bibinfo
  {title} {Cooper pair splitting by means of graphene quantum dots},}\ }\href
  {\doibase 10.1103/physrevlett.114.096602} {\bibfield  {journal} {\bibinfo
  {journal} {Physical Review Letters}\ }\textbf {\bibinfo {volume} {114}},\
  \bibinfo {pages} {096602} (\bibinfo {year} {2015})}\BibitemShut {NoStop}%
\bibitem [{\citenamefont {Borzenets}\ \emph {et~al.}(2016)\citenamefont
  {Borzenets}, \citenamefont {Shimazaki}, \citenamefont {Jones}, \citenamefont
  {Craciun}, \citenamefont {Russo}, \citenamefont {Yamamoto},\ and\
  \citenamefont {Tarucha}}]{Borzenets2016}%
  \BibitemOpen
  \bibfield  {author} {\bibinfo {author} {\bibfnamefont {I.~V.}\ \bibnamefont
  {Borzenets}}, \bibinfo {author} {\bibfnamefont {Y.}~\bibnamefont
  {Shimazaki}}, \bibinfo {author} {\bibfnamefont {G.~F.}\ \bibnamefont
  {Jones}}, \bibinfo {author} {\bibfnamefont {M.~F.}\ \bibnamefont {Craciun}},
  \bibinfo {author} {\bibfnamefont {S.}~\bibnamefont {Russo}}, \bibinfo
  {author} {\bibfnamefont {M.}~\bibnamefont {Yamamoto}}, \ and\ \bibinfo
  {author} {\bibfnamefont {S.}~\bibnamefont {Tarucha}},\ }\bibfield  {title}
  {\enquote {\bibinfo {title} {High efficiency {CVD} graphene-lead (pb) cooper
  pair splitter},}\ }\href {\doibase 10.1038/srep23051} {\bibfield  {journal}
  {\bibinfo  {journal} {Scientific Reports}\ }\textbf {\bibinfo {volume} {6}},\
  \bibinfo {pages} {23051} (\bibinfo {year} {2016})}\BibitemShut {NoStop}%
\bibitem [{\citenamefont {Fasth}\ \emph {et~al.}(2005)\citenamefont {Fasth},
  \citenamefont {Fuhrer}, \citenamefont {Björk},\ and\ \citenamefont
  {Samuelson}}]{Fasth2005}%
  \BibitemOpen
  \bibfield  {author} {\bibinfo {author} {\bibfnamefont {C.}~\bibnamefont
  {Fasth}}, \bibinfo {author} {\bibfnamefont {A.}~\bibnamefont {Fuhrer}},
  \bibinfo {author} {\bibfnamefont {M.~T.}\ \bibnamefont {Björk}}, \ and\
  \bibinfo {author} {\bibfnamefont {L.}~\bibnamefont {Samuelson}},\ }\bibfield
  {title} {\enquote {\bibinfo {title} {Tunable double quantum dots in inas
  nanowires defined by local gate electrodes},}\ }\href {\doibase
  10.1021/nl050850i} {\bibfield  {journal} {\bibinfo  {journal} {Nano Letters}\
  }\textbf {\bibinfo {volume} {5}},\ \bibinfo {pages} {1487--1490} (\bibinfo
  {year} {2005})}\BibitemShut {NoStop}%
\bibitem [{\citenamefont {Gramich}, \citenamefont {Baumgartner},\ and\
  \citenamefont {Sch{\"o}nenberger}(2015)}]{Gramich2015}%
  \BibitemOpen
  \bibfield  {author} {\bibinfo {author} {\bibfnamefont {J.}~\bibnamefont
  {Gramich}}, \bibinfo {author} {\bibfnamefont {A.}~\bibnamefont
  {Baumgartner}}, \ and\ \bibinfo {author} {\bibfnamefont {C.}~\bibnamefont
  {Sch{\"o}nenberger}},\ }\bibfield  {title} {\enquote {\bibinfo {title}
  {Resonant and inelastic andreev tunneling observed on a carbon nanotube
  quantum dot},}\ }\href {\doibase 10.1103/physrevlett.115.216801} {\bibfield
  {journal} {\bibinfo  {journal} {Physical Review Letters}\ }\textbf {\bibinfo
  {volume} {115}},\ \bibinfo {pages} {216801} (\bibinfo {year}
  {2015})}\BibitemShut {NoStop}%
\bibitem [{\citenamefont {Suyatin}\ \emph {et~al.}(2007)\citenamefont
  {Suyatin}, \citenamefont {Thelander}, \citenamefont {Björk}, \citenamefont
  {Maximov},\ and\ \citenamefont {Samuelson}}]{Suyatin2007}%
  \BibitemOpen
  \bibfield  {author} {\bibinfo {author} {\bibfnamefont {D.~B.}\ \bibnamefont
  {Suyatin}}, \bibinfo {author} {\bibfnamefont {C.}~\bibnamefont {Thelander}},
  \bibinfo {author} {\bibfnamefont {M.~T.}\ \bibnamefont {Björk}}, \bibinfo
  {author} {\bibfnamefont {I.}~\bibnamefont {Maximov}}, \ and\ \bibinfo
  {author} {\bibfnamefont {L.}~\bibnamefont {Samuelson}},\ }\bibfield  {title}
  {\enquote {\bibinfo {title} {Sulfur passivation for ohmic contact formation
  to {InAs} nanowires},}\ }\href {\doibase 10.1088/0957-4484/18/10/105307}
  {\bibfield  {journal} {\bibinfo  {journal} {Nanotechnology}\ }\textbf
  {\bibinfo {volume} {18}},\ \bibinfo {pages} {105307} (\bibinfo {year}
  {2007})}\BibitemShut {NoStop}%
\bibitem [{\citenamefont {Baba}\ \emph {et~al.}(2017)\citenamefont {Baba},
  \citenamefont {Matsuo}, \citenamefont {Kamata}, \citenamefont {Deacon},
  \citenamefont {Oiwa}, \citenamefont {Li}, \citenamefont {Jeppesen},
  \citenamefont {Samuelson}, \citenamefont {Xu},\ and\ \citenamefont
  {Tarucha}}]{Baba2017}%
  \BibitemOpen
  \bibfield  {author} {\bibinfo {author} {\bibfnamefont {S.}~\bibnamefont
  {Baba}}, \bibinfo {author} {\bibfnamefont {S.}~\bibnamefont {Matsuo}},
  \bibinfo {author} {\bibfnamefont {H.}~\bibnamefont {Kamata}}, \bibinfo
  {author} {\bibfnamefont {R.~S.}\ \bibnamefont {Deacon}}, \bibinfo {author}
  {\bibfnamefont {A.}~\bibnamefont {Oiwa}}, \bibinfo {author} {\bibfnamefont
  {K.}~\bibnamefont {Li}}, \bibinfo {author} {\bibfnamefont {S.}~\bibnamefont
  {Jeppesen}}, \bibinfo {author} {\bibfnamefont {L.}~\bibnamefont {Samuelson}},
  \bibinfo {author} {\bibfnamefont {H.~Q.}\ \bibnamefont {Xu}}, \ and\ \bibinfo
  {author} {\bibfnamefont {S.}~\bibnamefont {Tarucha}},\ }\bibfield  {title}
  {\enquote {\bibinfo {title} {Gate tunable parallel double quantum dots in
  {InAs} double-nanowire devices},}\ }\href {\doibase 10.1063/1.4997646}
  {\bibfield  {journal} {\bibinfo  {journal} {Applied Physics Letters}\
  }\textbf {\bibinfo {volume} {111}},\ \bibinfo {pages} {233513} (\bibinfo
  {year} {2017})}\BibitemShut {NoStop}%
\bibitem [{\citenamefont {Pillet}\ \emph {et~al.}(2010)\citenamefont {Pillet},
  \citenamefont {Quay}, \citenamefont {Morfin}, \citenamefont {Bena},
  \citenamefont {Yeyati},\ and\ \citenamefont {Joyez}}]{Pillet2010}%
  \BibitemOpen
  \bibfield  {author} {\bibinfo {author} {\bibfnamefont {J.-D.}\ \bibnamefont
  {Pillet}}, \bibinfo {author} {\bibfnamefont {C.~H.~L.}\ \bibnamefont {Quay}},
  \bibinfo {author} {\bibfnamefont {P.}~\bibnamefont {Morfin}}, \bibinfo
  {author} {\bibfnamefont {C.}~\bibnamefont {Bena}}, \bibinfo {author}
  {\bibfnamefont {A.~L.}\ \bibnamefont {Yeyati}}, \ and\ \bibinfo {author}
  {\bibfnamefont {P.}~\bibnamefont {Joyez}},\ }\bibfield  {title} {\enquote
  {\bibinfo {title} {Andreev bound states in supercurrent-carrying carbon
  nanotubes revealed},}\ }\href {\doibase 10.1038/nphys1811} {\bibfield
  {journal} {\bibinfo  {journal} {Nature Physics}\ }\textbf {\bibinfo {volume}
  {6}},\ \bibinfo {pages} {965--969} (\bibinfo {year} {2010})}\BibitemShut
  {NoStop}%
\bibitem [{\citenamefont {Dirks}\ \emph {et~al.}(2011)\citenamefont {Dirks},
  \citenamefont {Hughes}, \citenamefont {Lal}, \citenamefont {Uchoa},
  \citenamefont {Chen}, \citenamefont {Chialvo}, \citenamefont {Goldbart},\
  and\ \citenamefont {Mason}}]{Dirks2011}%
  \BibitemOpen
  \bibfield  {author} {\bibinfo {author} {\bibfnamefont {T.}~\bibnamefont
  {Dirks}}, \bibinfo {author} {\bibfnamefont {T.~L.}\ \bibnamefont {Hughes}},
  \bibinfo {author} {\bibfnamefont {S.}~\bibnamefont {Lal}}, \bibinfo {author}
  {\bibfnamefont {B.}~\bibnamefont {Uchoa}}, \bibinfo {author} {\bibfnamefont
  {Y.-F.}\ \bibnamefont {Chen}}, \bibinfo {author} {\bibfnamefont
  {C.}~\bibnamefont {Chialvo}}, \bibinfo {author} {\bibfnamefont {P.~M.}\
  \bibnamefont {Goldbart}}, \ and\ \bibinfo {author} {\bibfnamefont
  {N.}~\bibnamefont {Mason}},\ }\bibfield  {title} {\enquote {\bibinfo {title}
  {Transport through andreev bound states in a graphene quantum dot},}\ }\href
  {\doibase 10.1038/nphys1911} {\bibfield  {journal} {\bibinfo  {journal}
  {Nature Physics}\ }\textbf {\bibinfo {volume} {7}},\ \bibinfo {pages}
  {386--390} (\bibinfo {year} {2011})}\BibitemShut {NoStop}%
\bibitem [{\citenamefont {Gramich}, \citenamefont {Baumgartner},\ and\
  \citenamefont {Sch{\"o}nenberger}(2017)}]{Gramich2017}%
  \BibitemOpen
  \bibfield  {author} {\bibinfo {author} {\bibfnamefont {J.}~\bibnamefont
  {Gramich}}, \bibinfo {author} {\bibfnamefont {A.}~\bibnamefont
  {Baumgartner}}, \ and\ \bibinfo {author} {\bibfnamefont {C.}~\bibnamefont
  {Sch{\"o}nenberger}},\ }\bibfield  {title} {\enquote {\bibinfo {title}
  {Andreev bound states probed in three-terminal quantum dots},}\ }\href
  {\doibase 10.1103/physrevb.96.195418} {\bibfield  {journal} {\bibinfo
  {journal} {Physical Review B}\ }\textbf {\bibinfo {volume} {96}},\ \bibinfo
  {pages} {195418} (\bibinfo {year} {2017})}\BibitemShut {NoStop}%
\bibitem [{\citenamefont {Jellinggaard}\ \emph {et~al.}(2016)\citenamefont
  {Jellinggaard}, \citenamefont {Grove-Rasmussen}, \citenamefont {Madsen},\
  and\ \citenamefont {Nyg{\aa}rd}}]{Jellinggaard2016}%
  \BibitemOpen
  \bibfield  {author} {\bibinfo {author} {\bibfnamefont {A.}~\bibnamefont
  {Jellinggaard}}, \bibinfo {author} {\bibfnamefont {K.}~\bibnamefont
  {Grove-Rasmussen}}, \bibinfo {author} {\bibfnamefont {M.~H.}\ \bibnamefont
  {Madsen}}, \ and\ \bibinfo {author} {\bibfnamefont {J.}~\bibnamefont
  {Nyg{\aa}rd}},\ }\bibfield  {title} {\enquote {\bibinfo {title} {Tuning
  yu-shiba-rusinov states in a quantum dot},}\ }\href {\doibase
  10.1103/physrevb.94.064520} {\bibfield  {journal} {\bibinfo  {journal}
  {Physical Review B}\ }\textbf {\bibinfo {volume} {94}},\ \bibinfo {pages}
  {064520} (\bibinfo {year} {2016})}\BibitemShut {NoStop}%
\end{thebibliography}%
\end{document}